\documentclass[conference]{IEEEtran}
\IEEEoverridecommandlockouts

\usepackage{cite}
\usepackage{amsmath,amssymb,amsfonts}
\usepackage{algorithmic}
\usepackage{graphicx}
\usepackage{subfig}
\usepackage{textcomp}
\usepackage{xcolor}
\def\BibTeX{{\rm B\kern-.05em{\sc i\kern-.025em b}\kern-.08em
    T\kern-.1667em\lower.7ex\hbox{E}\kern-.125emX}}
\usepackage{caption}

\usepackage{fancyhdr}
\usepackage{etoolbox}

\fancypagestyle{firstpage}{
    \fancyhf{}
    \fancyhead[L]{\small
    This is the author’s accepted manuscript of a paper accepted at the 2025 IEEE International Conference on Biomedical Engineering, Computer and Information Technology for Health (BECITHCON). The final published version is available via IEEE Xplore.\\ DOI: https://doi.org/10.1109/BECITHCON69222.2025.11503952
    }

}

\begin{document}

\title{Brain Tumor Classification in MRI Images: A Computationally Efficient Convolutional Neural Network\\
}

\author{
\IEEEauthorblockN{Md Fahimul Kabir Chowdhury\textsuperscript{1} and Jannatul Ferdous\textsuperscript{2}}
\IEEEauthorblockA{
\textsuperscript{1}\textit{Department of Computer Science and Engineering, University of North Texas, USA} \\
\textsuperscript{2}\textit{Department of Electrical and Electronic Engineering, International Islamic University Chittagong, Bangladesh} \\
mdfahimulkabirchowdhury@my.unt.edu, ET223224@ugrad.iiuc.ac.bd
}
}
\maketitle
\thispagestyle{firstpage}  

\begin{abstract}
Improving patient outcomes depends on the prompt and accurate diagnosis of brain tumors, but manual MRI scan analysis is still time-consuming and unreliable. Although deep learning has shown promise, many of the models that are now in use are computationally intensive and have difficulty handling the intrinsic complexity and variety of different types of brain tumors. In this work, we propose a lightweight yet high-performing Convolutional Neural Network (CNN) for multi-class brain tumor classification, employing MRI images to target gliomas, meningiomas, pituitary tumors, and healthy (no tumor) instances. The model was rigorously evaluated on two publicly accessible datasets from Figshare and Kaggle. Leveraging efficient feature extraction and optimized training strategies, our CNN achieved classification accuracies of 99.03\% and 99.28\%, along with ROC scores of 99.88\% and 99.94\% on Dataset 1 and Dataset 2, respectively-all while utilizing significantly fewer parameters than popular pre-trained architectures. In contrast to cutting-edge models like DenseNet201, MobileNetV2, VGG19, Xception, InceptionV3, and ResNet50, our approach consistently demonstrated superior performance with reduced computational overhead. These findings highlight the potential of the proposed model as a practical and reliable diagnostic aid in clinical environments.
\end{abstract}


\section{Introduction}
Neurological disorders affect individuals across all age groups, encompassing a wide spectrum of conditions that present significant public health challenges. Brain tumors, in particular, manifest through diverse symptoms such as persistent headaches, cognitive decline, or severe neurological impairments, which change depending on the kind, size, and location of the tumor \cite{munir2019cancer,thakkar2020rehabilitation}. Advances in non-invasive approaches, especially through computer-aided diagnosis (CAD) systems, have enhanced the prospects for early detection. Such timely diagnosis is crucial for improving treatment outcomes in life-threatening conditions like brain cancer, where survival rates remain critically low.

\begin{figure}[t]
    \centering
    \subfloat[Glioma]{%
        \includegraphics[height=2cm, width=0.11\textwidth]{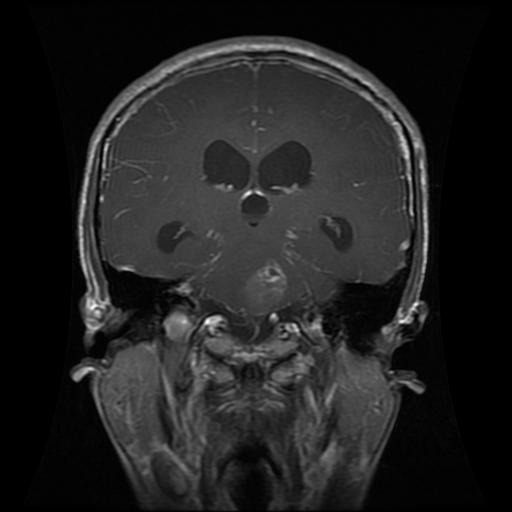}%
        \label{fig:glioma}
    }
    \hfill
    \subfloat[Meningioma]{%
        \includegraphics[height=2cm, width=0.11\textwidth]{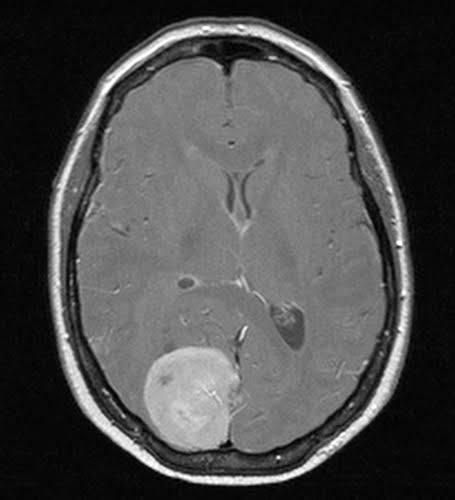}%
        \label{fig:meningioma}
    }
    \hfill
    \subfloat[Pituitary]{%
        \includegraphics[height=2cm, width=0.11\textwidth]{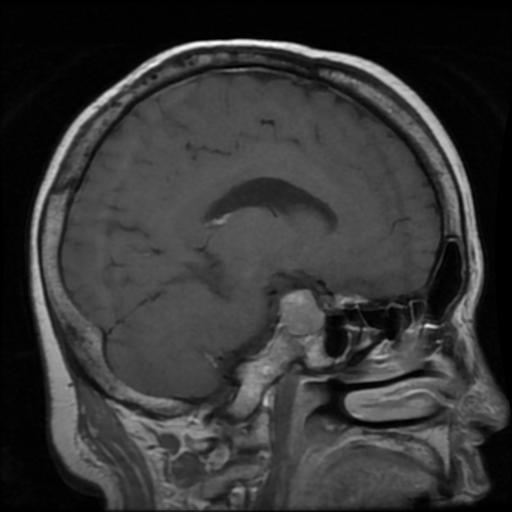}%
        \label{fig:pituitary}
    }
    \hfill
    \subfloat[No Tumor]{%
        \includegraphics[height=2cm, width=0.11\textwidth]{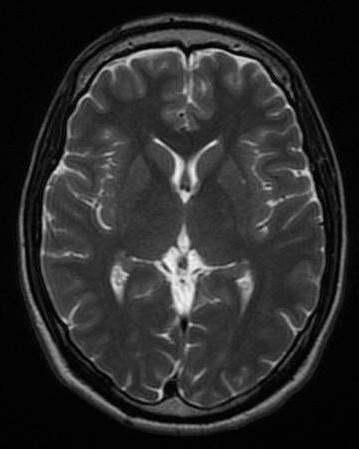}%
        \label{fig:notumor}
    }

    \caption{Representative images from each brain tumor class in the dataset.}
    \label{fig:sample_classes}
\end{figure}

Despite notable progress in medical imaging, brain tumor diagnosis continues to be a challenging and time-intensive task, often demanding expert evaluation and prolonged waiting periods \cite{thakkar2020rehabilitation}. Figure \ref{fig:sample_classes} illustrates why manual interpretation is challenging by displaying a variety of sample photos of various tumor kinds.

Deep learning, particularly through CNNs, has accomplished impressive results in picture classification, including uses in medical imaging \cite{munir2019cancer}. From unprocessed photos, CNNs may automatically extract hierarchical features, which makes them effective for segmentation, feature learning, and classification tasks. This field has advanced since transfer learning was introduced since it makes pre-trained models applicable to domain-specific issues \cite{ozkaraca2023multiple,chowdhury2022development}. This strategy reduces both training time and computational requirements while maintaining strong performance \cite{srinivas2022deep}.

Nonetheless, many of these approaches are evaluated on relatively small datasets, as access to large medical image repositories remains limited. Moreover, a significant portion of earlier studies relied on pre-trained CNN models originally developed for natural image datasets. While these models can be fine-tuned for brain tumor analysis, they are not inherently designed to capture domain-specific tumor patterns, which restricts their effectiveness in brain tumor diagnosis. Current research frequently uses transfer learning using pre-trained models or deep CNN architectures, which are memory-intensive and computationally costly. For real-time diagnosis, where quick results are essential, such methods might not be appropriate.  Although there are various lightweight CNN variations, many of them do not include multi-scale feature extraction, which is essential for collecting minor tumor properties. Additionally, comprehensive investigations that strike a balance between interpretability and accuracy are scarce, an essential aspect of clinical decision-making.

In this work, we design a lightweight and efficient CNN-based method for classifying brain tumors from MRI scans. Common metrics such as F1-score, precision, recall, accuracy, and AUC (PR/ROC) are used to assess model performance. Two different datasets comprising images of Glioma (GL), Meningioma (ME), Pituitary (PI) tumors, and a category of No Tumor (NT) is constructed from publicly available repositories. The proposed approach is benchmarked against widely used pre-trained models such as MobileNetV2, DenseNet201, VGG19, Xception, InceptionV3, and ResNet50, as well as against existing conventional methods.

\vspace{1em}
\noindent The following are this work's main contributions:
\begin{itemize}
\item A lightweight CNN architecture is introduced to achieve accurate classification of gliomas, meningiomas, and pituitary tumors are the three main forms of brain tumors.

\item We perform a comparative analysis to identify the types of brain tumor more accurately between several cutting-edge pre-trained architectures and the suggested CNN.

\item We reduce computational overhead by avoiding heavy preprocessing, deep network structures, extensive data augmentation, and reliance on pre-trained models.

\item We evaluate model performance using an unbiased cross-validation (CV) strategy, and assess performance using metrics including F1-score, recall, accuracy, and precision to identify the best-performing configuration for predicting depression treatment outcomes.

\end{itemize}

\section{Related Work}

\subsection{Convolutional Neural Network}


Islam et al. \cite{islam2024improved}, have enhanced brain tumor diagnosis through MRI feature extraction and deep learning methods. Using three merged datasets, 2D CNN, CNN-LSTM, and ensemble models achieved up to 98.82\% accuracy, 99\% precision, and 99\% recall, with the ensemble outperforming individual models. Aamir et al. \cite{aamir2024brain}, proposed a CNN with hyperparameters adjusted using three publicly findable Kaggle MRI datasets, optimizing factors like batch size, learning rate, and filter size. The model achieved ~97\% accuracy, outperforming existing methods and enhancing generalization for reliable clinical diagnosis.

This \cite{martinez2024brain}, study applied CNN-based classifiers on MRI scans, achieving 97.5\% accuracy, 99.2\% sensitivity, and 98.2\% binary accuracy, outperforming prior works. In \cite{agrawal2024multifenet}, introduced MultiFeNet, a CNN-based model with multi-scale feature extraction from the Figshare dataset, evaluated on 3,064 MRI scans (meningioma, glioma, pituitary). The model achieved 96.4\% accuracy, precision, sensitivity, and F1-score. Khan et al.\cite{khan2025lead}, proposed LEAD-CNN, with a dimension reduction block(lightweight), evaluated on 7,023 MRI images (three tumor types + normal) from Kaggle. The model achieved 98.70\% accuracy, F1-score 98.62\% with recall 98.6\%, and precision 98.65\%. Batool et al. \cite{batool2025lightweight}, introduced a lightweight Multi-path CNN (M-CNN) with an optimal feature selection module, compared against CNN, D-CNN, AlexNet, ResNet, and Inception V3. The M-CNN demonstrated better classification performance and computational efficiency while lowering overfitting, achieving 92.25\% accuracy with all features and 96.03\% with selected features.

\subsection{Hybrid and Transfer Learning(TF)}


Sathya et al. \cite{sathya2024employing}, suggested a model based on Xception that includes batch normalization, dropout, and customized dense layers, using transfer learning on meningioma, glioma, and pituitary tumor MRI data. The model achieved 98.04\% accuracy and 96\% precision and recall. Incir et al. \cite{incir2024improving}, evaluated six transfer learning models (ResNet-50, MobileNet, VGG16, Inception-V3, DenseNet-121, EfficientNetV2-M) on a public MRI dataset with data augmentation, finding EfficientNetV2-M achieved 98.01\% accuracy. Accuracy was further increased to 98.41\% using a concatenation-based model that combined EfficientNetV2-M + Inception-V3. Shaha et al. \cite{shaha2025mri}, compared deep learning (CNNs: InceptionV3, VGG16/19, MobileNetV2, ResNet50) with SVM using transfer learning and GLCM features on MRI scans. CNNs outperformed SVM, with InceptionV3 achieving 94.12\% accuracy and highest recall, demonstrating superior scalability, multi-scale feature extraction, and generalization, while SVM was limited on larger datasets.

Ticku et al. \cite{ticku2025advancing}, suggested a Quantum CNN (QCNN) with four-qubit convolution layers and quantum embedding, developed with PennyLane and TensorFlow Quantum, tested on 3,000+ MRI scans. The QCNN achieved 92.13\% accuracy, comparable to ResNet50, while reducing training time from 64.95 s to 1.1 s, highlighting its potential for faster, efficient real-time diagnostics. In \cite{shinde2025high}, proposed a Parallel Quantum Dilated CNN (PQDCNN) with Map-Reduce, using FLICM clustering, Medav filtering, TransBTSV2 segmentation, and feature extraction. The model achieved 91.52\% accuracy, 91.69\% sensitivity, and 92.26\% specificity, demonstrating effective and high-performance brain tumor detection.

Ilani et al. \cite{ilani2025t1}, applied U-Net segmentation and CNNs (Inception-V3, EfficientNetB4, VGG19) with transfer learning on MRI scans. U-Net achieved 98.56\% accuracy, and 96.01\% cross-dataset accuracy. In \cite{mzoughi2025vision}, a Vision Transformer (ViT) architecture, T1-weighted contrast-enhanced MRI was used to train XAI algorithms (Grad-CAM, LIME, and SHAP) for multi-class tumor identification. Experimental results showed ViT achieved 91.61\% accuracy compared to 83.37\% for CNN. In \cite{gayathiri2025c}, A Convolutional Stacked Autoencoder Network (C-SAN), integrating CNN and DSAE, was proposed with preprocessing via NLM filter, segmentation using V-Net, and advanced feature extraction (MVM-LBP, DWT, HOG, LVP). Experimental results achieved 90.9\% accuracy, 95.8\% sensitivity, and 92.8\% specificity, showing C-SAN’s effectiveness for tumor detection.

\section{Methodology}

\begin{figure*}[t]
    \centering
    \includegraphics[width=\textwidth]{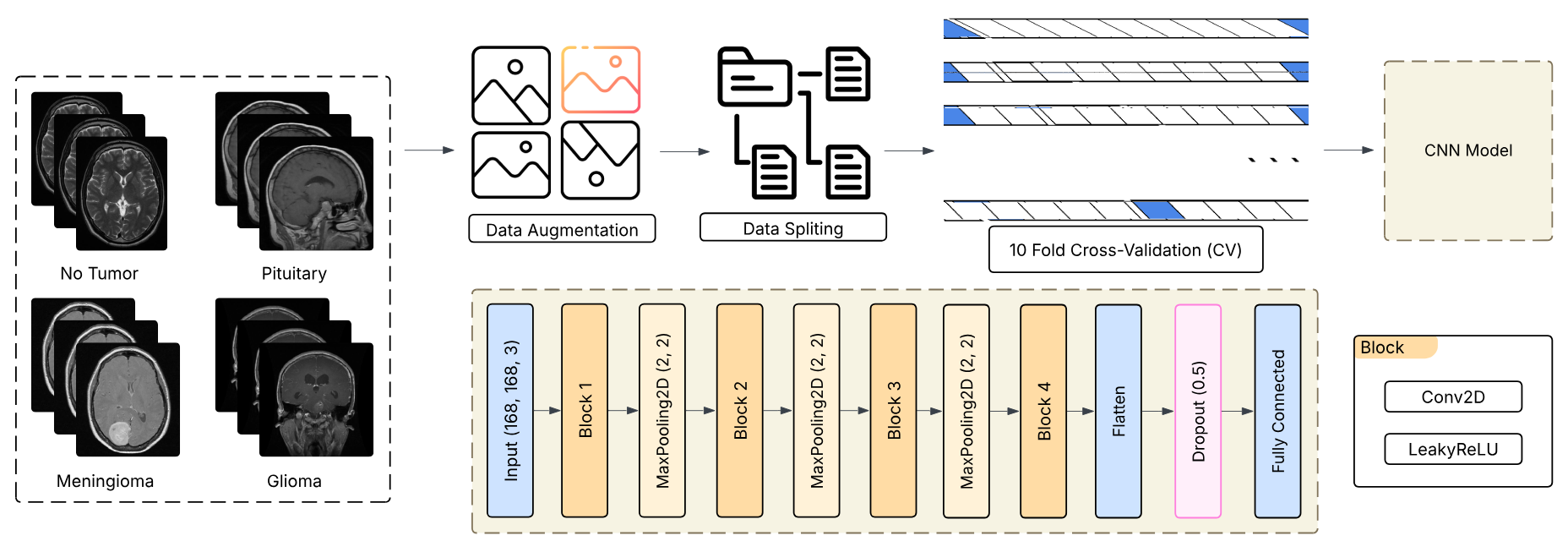}
    \caption{The process of data loading to propose a lightweight CNN architecture.}
    \label{blockD}
\end{figure*}

Our experiments are carried out on a powerful computer system that has two NVIDIA H100 NVL GPUs, each with 95,830 MB of memory, supported by CUDA 12.2. The machine’s substantial output computational capabilities enabled with efficient training, fine-tuning, and evaluation of all DL models. Figure~\ref{blockD} presents a high-level depiction of the architecture we propose.

\subsection{Data Description and Splitting}
In this work, two publicly findable datasets are utilized to evaluate the proposed CNN for brain tumor. The first dataset, obtained from Kaggle \cite{nickparvar2021dataset}, includes 7,023 brain MRI pictures categorized into four groups: meningioma, pituitary tumor, glioma, and healthy scans. From Figshare \cite{cheng2017dataset}, we were collected the second dataset which includes 233 patients provided 3,064 T1-weighted contrast-enhanced pictures. This dataset includes 930 pituitary tumor slices, 1,426 glioma slices, and 708 meningioma slices that were obtained from Nanfang Hospital and General Hospital in China.

To evaluate the performance of the suggested model, a stratified 10-fold cross-validation strategy we applied. To ensure reproducibility and prevent data leakage, we implement a custom script that systematically partitions the datasets and retrieves all image paths along with their corresponding class labels, storing them as paired entries. Using \texttt{KFold}, the dataset is randomly shuffled and divided into ten folds, with a fixed random seed to guarantee consistent splits. 

\subsection{Data Augmentation}

To enhance the robustness of the model and mitigate overfitting to the training set, data augmentation is applied. The augmentation strategy included horizontal flipping, small random rotations ($\pm 2^\circ$), brightness adjustment (0.8--1.2 scaling), minor zoom variations (95--105\%), and slight horizontal and vertical shifts (1\%). Both training and testing images are normalized by adjusting pixel values to fall between 0 and 1. For training, the augmented images are generated in batches of 32 with a target size of $168 \times 168$ pixels. For testing, only normalization was applied without augmentation to ensure unbiased evaluation.

\subsection{Proposed CNN Architecture}

The proposed network employs a series of convolutional layers as the core feature extraction mechanism. Convolutional layers process the input picture using a collection of learnable filters, capturing local spatial patterns such as edges, textures, and fine structures. When several layers are stacked, the model progressively learns more abstract and complex representations, transitioning from low-level features to tumor-specific characteristics.  

In our architecture, the first two convolutional blocks use 64 filters of size $3 \times 3$ with ``same'' padding, ensuring that spatial resolution is preserved. The subsequent two blocks employ 128 filters of the same kernel size, enabling the network to identify more intricate structures, such as irregular tumor boundaries and heterogeneous textures. The LeakyReLU activation function comes after each convolutional layer, which addresses the vanishing gradient problem and ensures non-linear feature learning. To progressively reduce the spatial dimensions while retaining discriminative features, max-pooling with a pool size of $2 \times 2$ is used after every convolutional block. This operation provides translational invariance and lowers computational cost, which is particularly advantageous for real-time diagnostic applications.  

The layered combination of convolution, non-linear activation, and pooling creates a hierarchical feature representation, where the shallow layers highlight minute details, whereas deeper layers record tumor morphology and contextual information critical for classification. A specific summary of this framework is shown in Fig. ~\ref{blockD}, and hyperparameters are listed in Table~\ref{Training_Hyperparameters}.

\begin{table}[t]
\centering
\caption{Training Hyperparameters Across Model Architectures}
\label{Training_Hyperparameters}
\begin{tabular}{|l|c|c|}
\hline
\textbf{Hyperparameters} & \textbf{pre-trained} & \textbf{Proposed CNN} \\
\hline
Input shape & (168, 168, 3) & (168, 168, 3) \\
Batch size & 32 & 32 \\
Dropout rate & 0.5 & 0.5 \\
Learning rate & 0.001 & 0.001 \\
Optimizer & AdamW & AdamW \\
Activation & LeakyReLU & LeakyReLU \\
Cross-Validation & 10-fold & 10-fold \\
Total params & 23,905,060 (Lowest) & 13,372,484 \\
Trainable params & 23,870,628 (Lowest) & 13,372,484 \\
\hline
\end{tabular}
\end{table}

\subsection{Implementation of Pre-Trained CNN Models}

To establish a fair comparison with the proposed CNN, several cutting-edge pre-trained models are implemented, including DenseNet201, MobileNetV2, VGG19, Xception, InceptionV3, and ResNet50. All models are initialized with ImageNet weights and configured with a fixed input shape of $168 \times 168 \times 3$. The final architecture for each model followed a consistent design to ensure comparability: with 1024 units of fully connected dense layer followed by a global average pooling layer, a LeakyReLU activation function, a softmax output layer for four-class classification and a dropout layer (rate = 0.5).  

Maintaining a uniform classification head across all pre-trained networks allowed us to isolate the effect of the feature extractors themselves, rather than differences in architectural choices.

\subsection{Cross Validation}

A 10-fold cross-validation approach is used to provide reliable performance assessments and lower the possibility of overfitting. In this approach, the datasets are partitioned into ten equal subsets. Nine subsets are utilized for training and one is applied for testing for each fold. The total performance is calculated by average the output over all folds, and this procedure is repeated repeatedly until each subset has been used as a test set precisely once.  

Because 10-fold cross-validation lowers the variance related to data splitting, it offers a more accurate estimate of model generalization than a single train-test split. This method also ensures that all samples contribute to both training and validation, consequently offering a comprehensive evaluation of the model's predictive ability.

\subsection{Model evaluation and validation}

Although numerous metrics exist to find out the performance of classification models, accuracy remains a core and commonly adopted criterion. In this work, we focus primarily on accuracy as the principal evaluation metric.

\begin{equation}
\text{Accuracy} = \frac{TP + TN}{TP + FP + TN + FN}
\end{equation}

where $TP$ and $TN$ indicates as true positives and true negatives, whereas $FP$ and $FN$ indicates as false positives and false negative.

We also utilize the confusion matrix, which offers a tabular breakdown of accurate and inaccurate forecasts across both classes. From this, we derive additional key metrics:

    \begin{equation}
    \text{Precision} = \frac{TP}{TP + FP}
    \end{equation}

    \begin{equation}
    \text{Recall} = \frac{TP}{TP + FN}
    \end{equation}

\begin{equation}
\text{F-score} = \frac{2}{3} \sum_{c=1}^{3} \frac{\text{Precision}_c \cdot \text{Recall}_c}{\text{Precision}_c + \text{Recall}_c}
\end{equation}

Additionally, we incorporate the Receiver Operating Characteristic (ROC) curve as a visual and quantitative tool to evaluate model performance. One important indicator of class separability is the Area Under the Curve (AUC). A perfect model yields an AUC of 1.0, indicating flawless discrimination between responder and non-responder classes.

\section{Results and Discussion}

\begin{figure*}[t]
    \centering
    \subfloat[MobileNetV2]{%
        \includegraphics[width=0.3\linewidth]{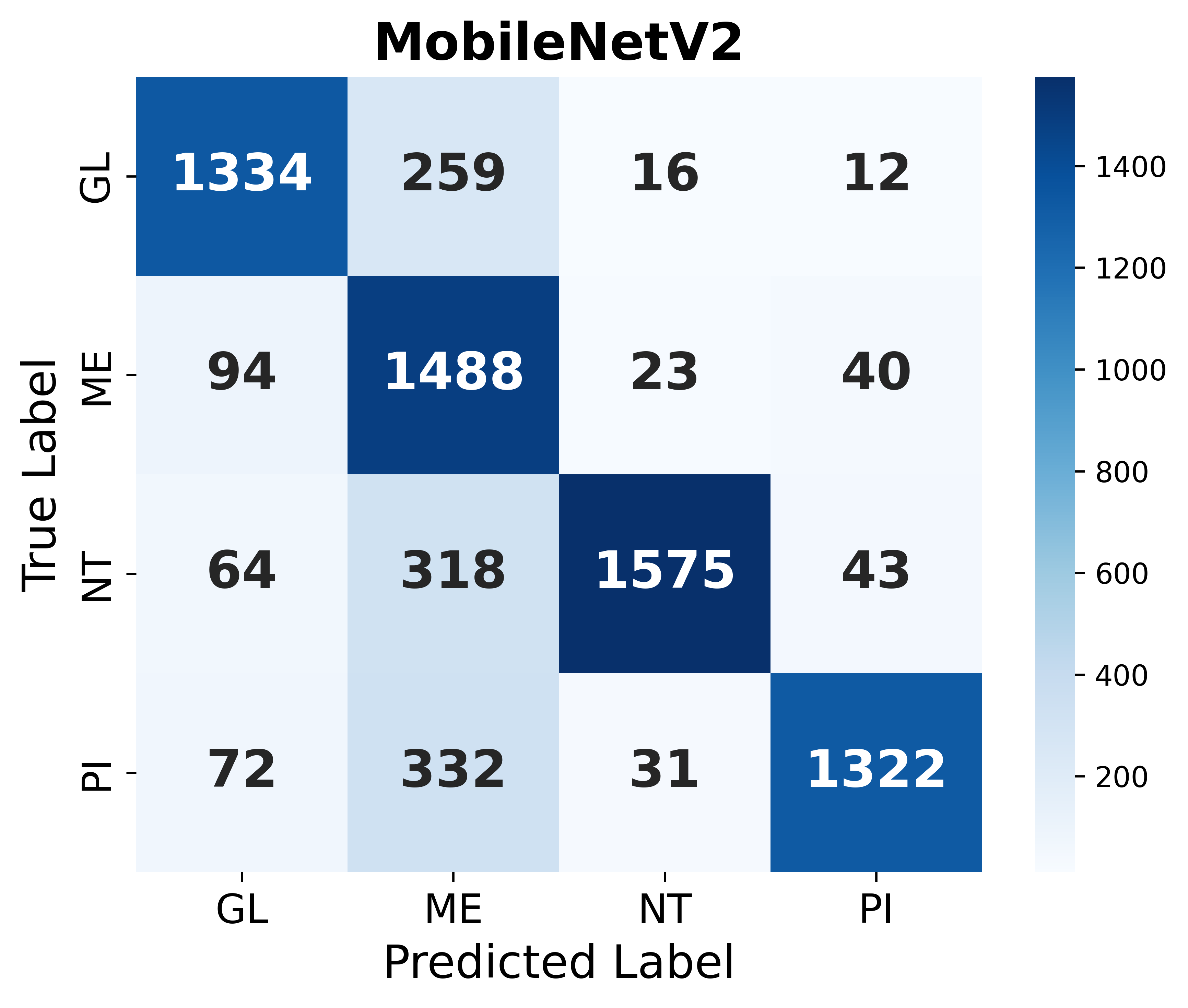}
    }
    \hfil
    \subfloat[DenseNet201]{%
        \includegraphics[width=0.3\linewidth]{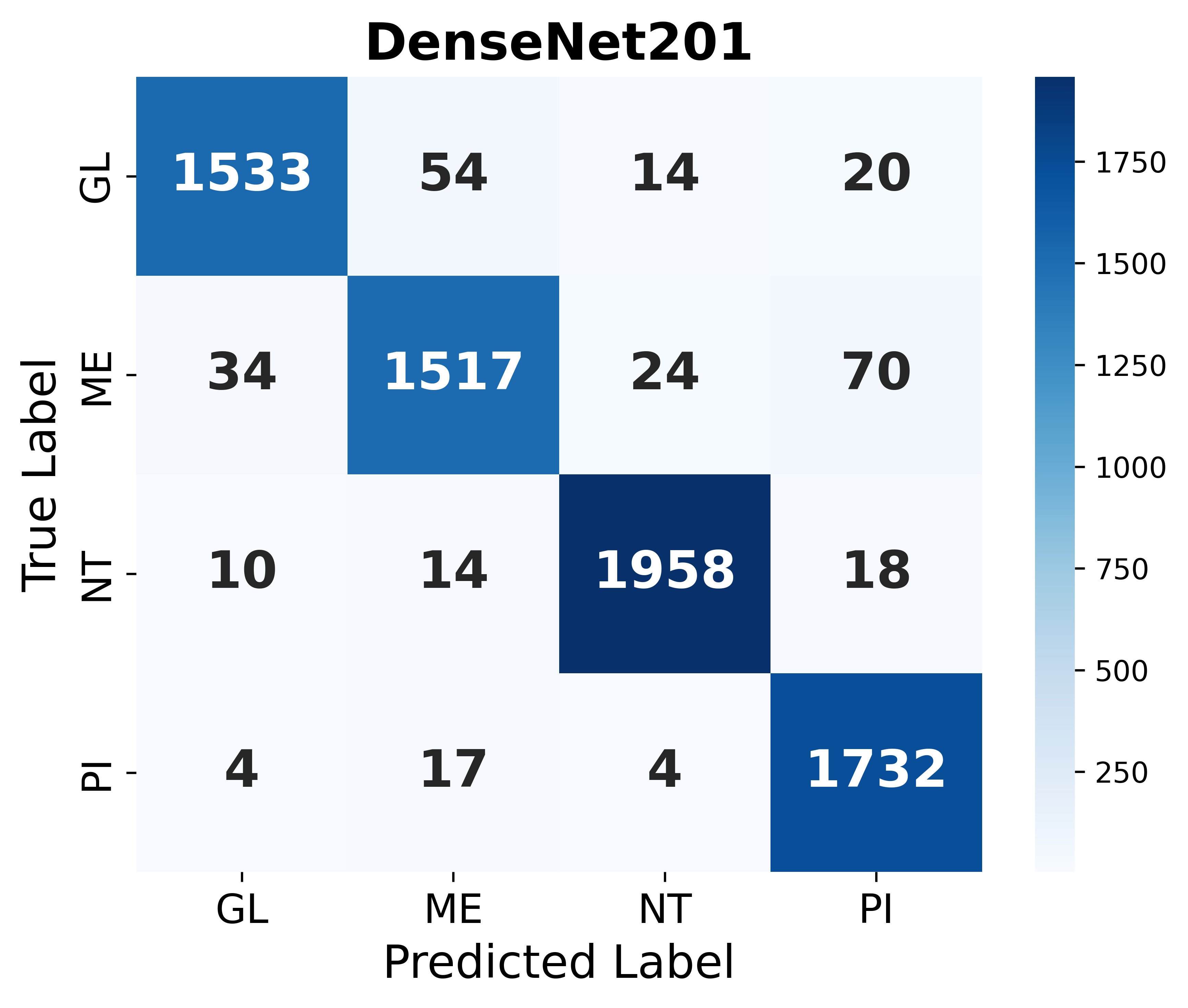}
    }
    \hfil
    \subfloat[Proposed CNN]{%
        \includegraphics[width=0.3\linewidth]{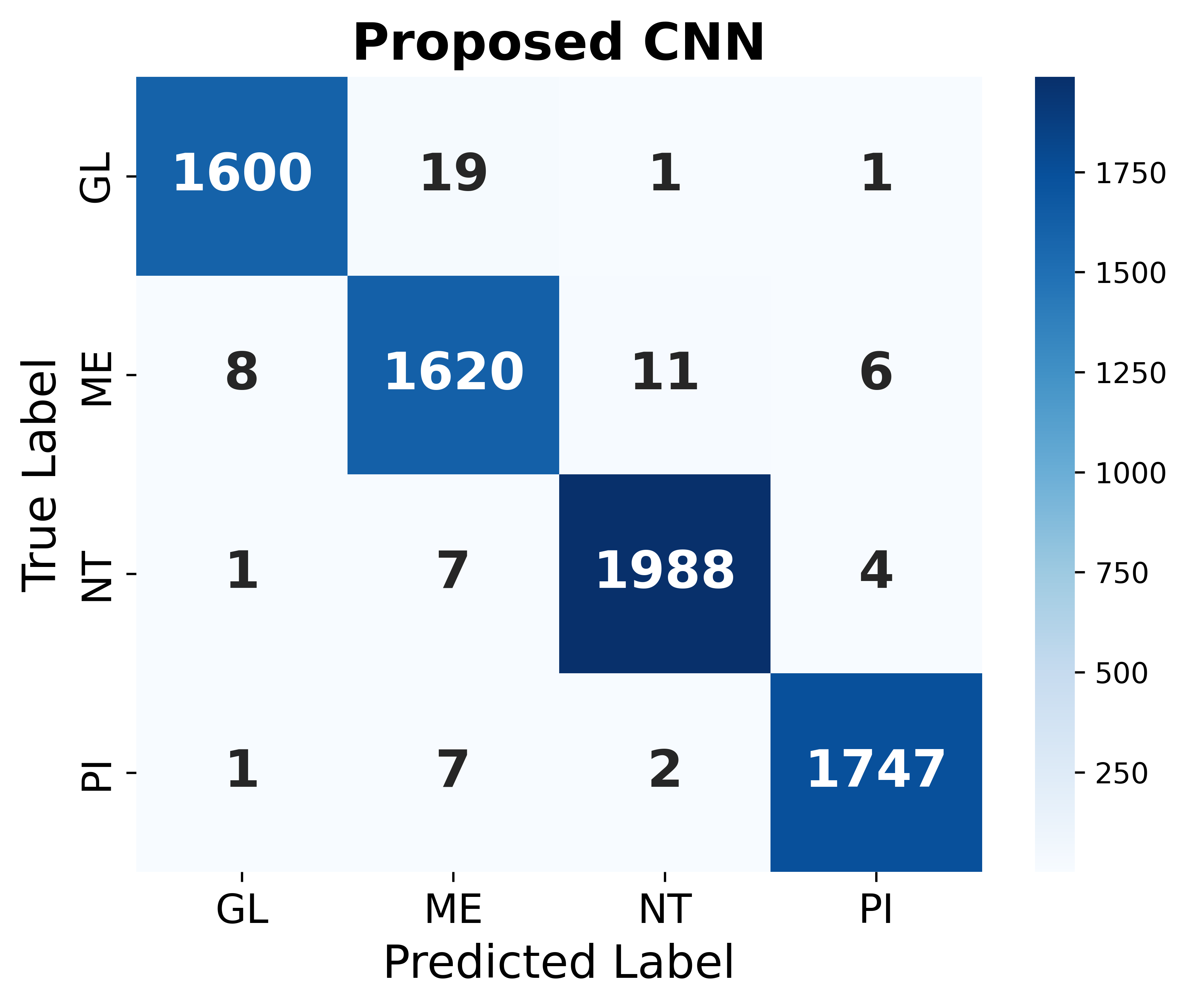}
    }
    \caption{Confusion matrix comparison of proposed CNN VS pre-trained models}
    \label{Fig:cm_comp}
\end{figure*}

In our first experiment, we contrast our suggested CNN with a number of state-of-the-art pre-trained models utilizing our first dataset. The findings presented in Table \ref{tab:cnn_results} shows that the proposed CNN model surpasses all other evaluated architectures in brain tumor classification across key metrics, including Precision, Recall, F1-score, Accuracy, and ROC. Specifically, the proposed CNN achieved a perfect 99\% Precision, Recall, and F1-score for all tumor categories, significantly surpassing the performance of pre-trained architectures such as VGG19, MobileNetV2, and DenseNet201. While Xception and InceptionV3 exhibited competitive performance, with accuracies of 98.31\% and 97.29\%, respectively, At 99.03\%, the suggested CNN had the best overall accuracy and an exceptional ROC score of 99.88\%. In Fig. \ref{Fig:cm_comp}, the Confusion Matrix comparison of randomly picked two pre-trained models vs our suggested CNN model.

\begin{table*}[t]
\centering
\caption{Comparing the suggested CNN's performance with pre-trained models on dataset 1.}
\label{tab:cnn_results}
\begin{tabular}{|l|c c c c|c c c c|c c c c|c|c|}
\hline
\textbf{Models} & \multicolumn{4}{c|}{\textbf{Precision}} & \multicolumn{4}{c|}{\textbf{Recall}} & \multicolumn{4}{c|}{\textbf{F1-score}} & \textbf{Accuracy (\%)} & \textbf{ROC (\%)} \\ \hline
 & GL & ME & NT & PI & GL & ME & NT & PI & GL & ME & NT & PI &  &  \\ \hline
VGG19        & 49 & 99 & 32 & 99 & 28 & 10 & 96 & 9  & 28 & 18 & 48 & 17 & 36.55 & 61.91 \\
MobileNetV2  & 85 & 62 & 96 & 93 & 82 & 90 & 79 & 75 & 84 & 74 & 86 & 83 & 81.43 & 94.36 \\
DenseNet201  & 97 & 95 & 98 & 94 & 95 & 92 & 98 & 99 & 96 & 93 & 98 & 96 & 95.97 & 99.66 \\
ResNet50     & 96 & 96 & 98 & 96 & 95 & 93 & 99 & 99 & 96 & 95 & 98 & 97 & 96.65 & 99.78 \\
InceptionV3  & 98 & 95 & 99 & 97 & 95 & 96 & 98 & 99 & 97 & 96 & 99 & 98 & 97.29 & 99.79 \\
Xception     & 98 & 97 & 99 & 99 & 98 & 97 & 100& 99 & 98 & 97 & 99 & 99 & 98.31 & 99.82 \\
\textbf{Proposed CNN} & \textbf{99} & \textbf{98} & \textbf{99} & \textbf{99} & \textbf{99} & \textbf{98} & \textbf{99} & \textbf{99} & \textbf{99} & \textbf{98} & \textbf{99} & \textbf{99} & \textbf{99.03} & \textbf{99.88} \\ \hline
\end{tabular}
\end{table*}

\begin{figure*}[t]
    \centering
    \subfloat[ResNet50]{%
        \includegraphics[width=0.3\linewidth]{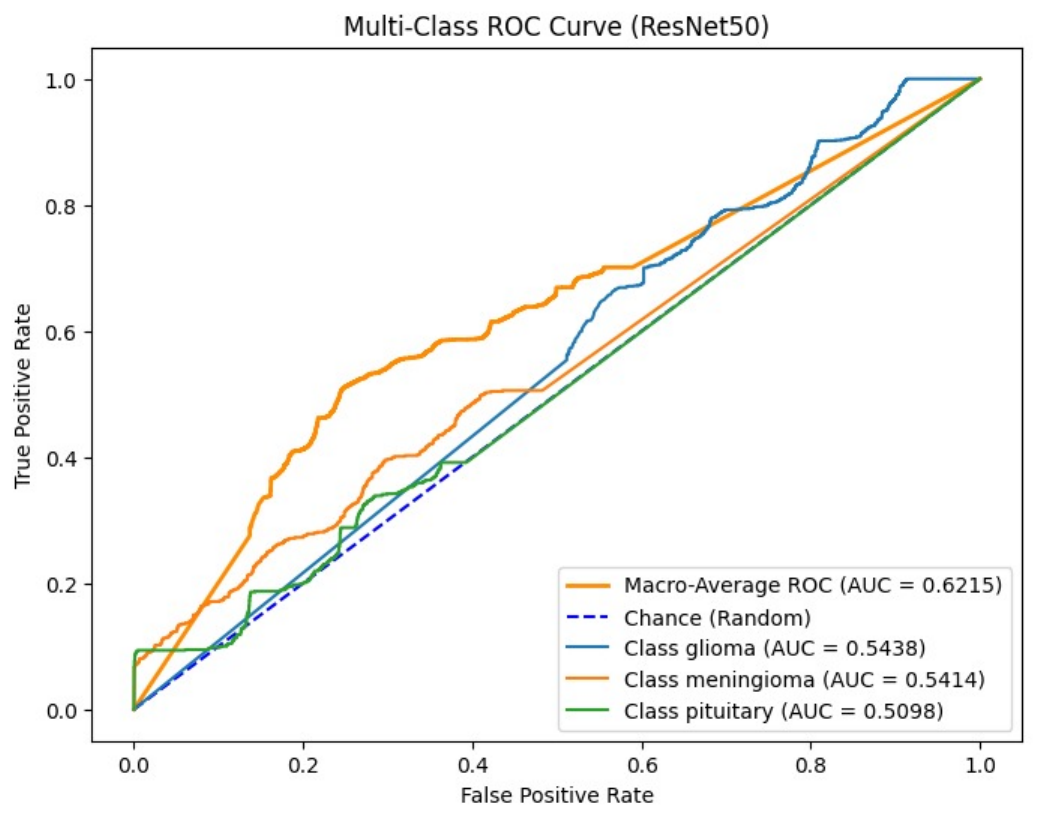}
    }
    \hfil
    \subfloat[InceptionV3]{%
        \includegraphics[width=0.3\linewidth]{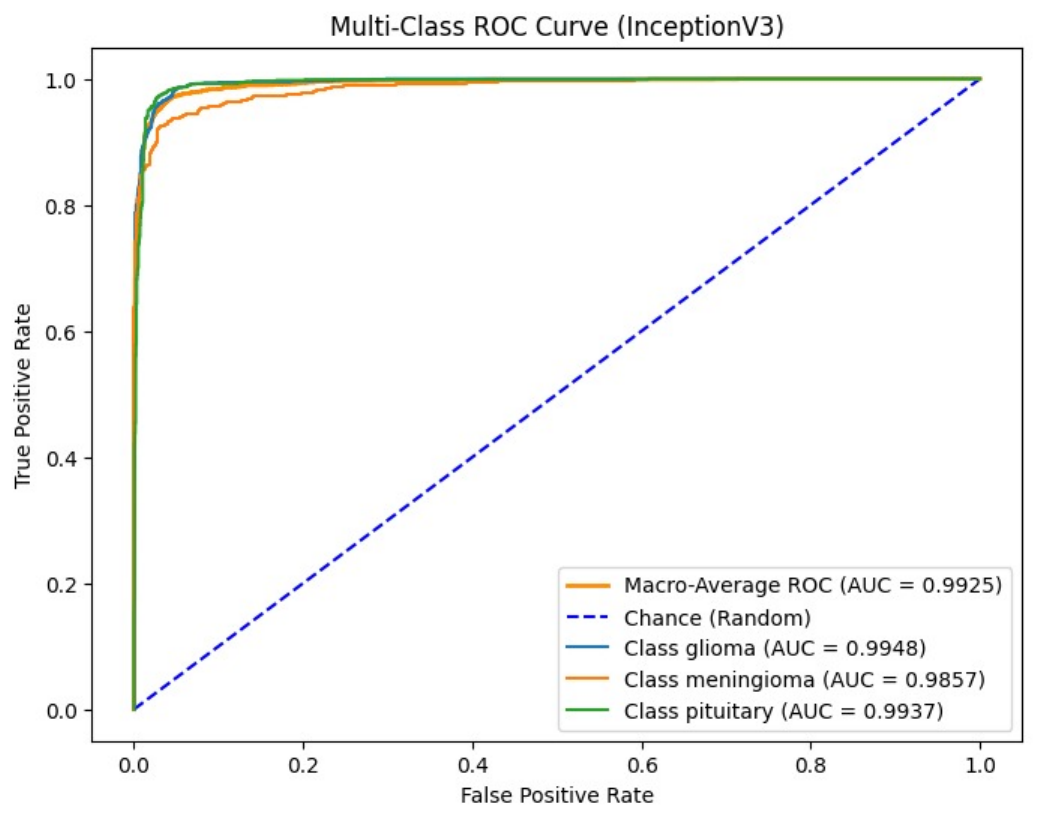}
    }
    \hfil
    \subfloat[Proposed CNN]{%
        \includegraphics[width=0.3\linewidth]{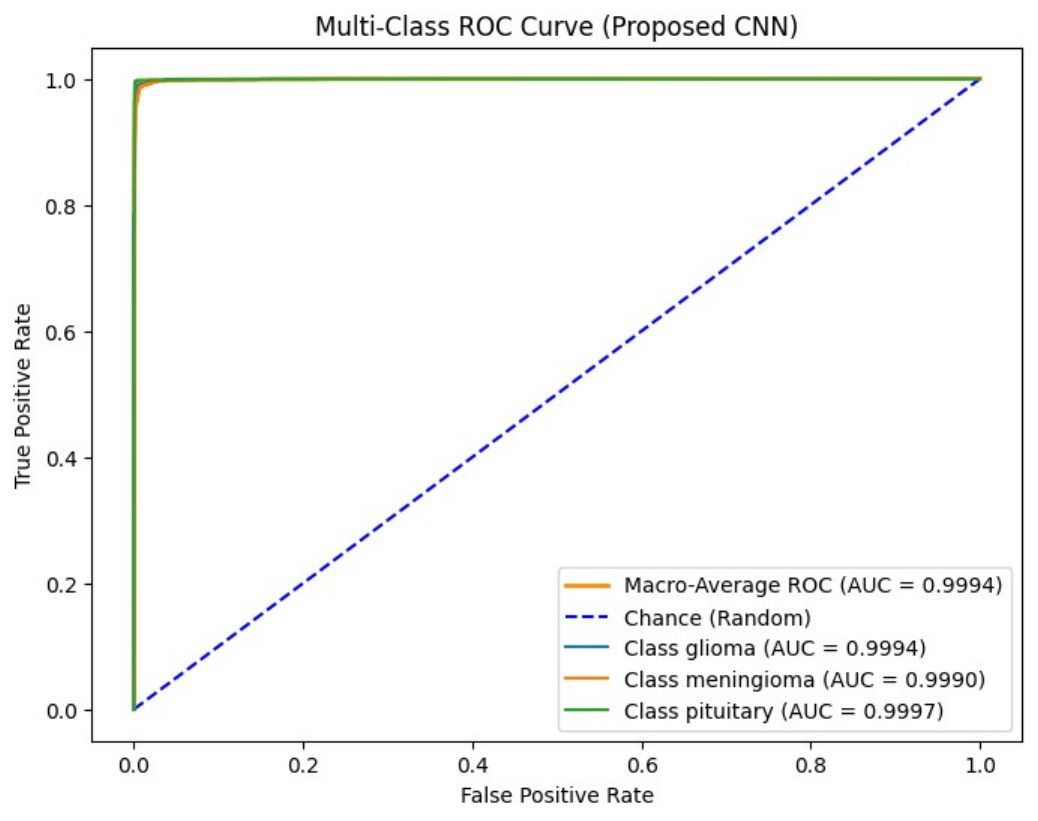}
    }
    \caption{ROC comparison of proposed CNN VS pre-trained models}
    \label{Fig:roc_comp}
\end{figure*}

In the second experiment shows in Table \ref{tab:cnn_results2}, we have used the second dataset to compare our suggested CNN's performance versus cutting-edge pre-trained models. Again in terms of all the important metics, the CNN performs better than any other models. It achieved perfect precision and recall (99\%) for all tumor types, and an F1-score of 100\% for Pituitary, 99\% for Glioma, and Meningioma. In terms of accuracy, the proposed CNN achieved 99.28\%, which is the highest among all models, with Xception being the second-best at 97.49\%. The ROC scores followed a similar trend, with the proposed CNN achieving an outstanding score of 99.94\%, surpassing Xception and InceptionV3, which achieved ROC values of 99.75\% and 99.25\%, accordingly. In Fig. \ref{Fig:roc_comp}, the ROC comparison of randomly picked two pre-trained models vs our proposed CNN model.

\begin{table*}[t]
\centering
\caption{Comparing the suggested CNN's performance with pre-trained models on dataset 2.}
\label{tab:cnn_results2}
\begin{tabular}{|l|c c c|c c c|c c c|c|c|}
\hline
\textbf{Models} & \multicolumn{3}{c|}{\textbf{Precision}} & \multicolumn{3}{c|}{\textbf{Recall}} & \multicolumn{3}{c|}{\textbf{F1-score}} & \textbf{Accuracy (\%)} & \textbf{ROC (\%)} \\ \hline
 & GL & ME & PI & GL & ME & PI & GL & ME & PI & & \\ \hline
ResNet50      & 49 & 100 & 87 & 100 & 6  & 17 & 66 & 12 & 9  & 50.78 & 62.15 \\
MobileNetV2   & 59 & 79  & 92 & 98  & 41 & 55 & 74 & 54 & 38 & 66.32 & 78.69 \\
VGG19         & 60 & 84  & 92 & 96  & 35 & 62 & 74 & 49 & 47 & 67 & 86.08 \\
InceptionV3   & 94 & 96  & 95 & 99  & 85 & 95 & 96 & 90 & 96 & 94.75 & 99.25 \\
DenseNet201   & 95 & 93  & 98 & 98  & 89 & 96 & 96 & 91 & 97 & 95.01 & 99.33 \\
Xception      & 98 & 96  & 98 & 99  & 95 & 98 & 98 & 96 & 97 & 97.49 & 99.75 \\
\textbf{Proposed CNN}  & \textbf{99} & \textbf{99}  & \textbf{100} & \textbf{99}  & \textbf{99} & \textbf{100} & \textbf{99} & \textbf{99} & \textbf{100} & \textbf{99.28} & \textbf{99.94} \\ \hline
\end{tabular}
\end{table*}

Despite its superior performance, the proposed CNN model requires fewer parameters compared to the larger pre-trained models shows in Table \ref{Training_Hyperparameters}, highlighting its efficiency. This suggests that the suggested architecture not only attains a high level of classification accuracy but also does so with a reduced computational load.

\begin{table}[t]
\centering
\caption{Comparative analysis with Proposed CNN on Dataset 1}
\label{tab:dataset1_accuracy_sorted}
\begin{tabular}{|l|c|c|}
\hline
\textbf{Study} & \textbf{Technique} & \textbf{Accuracy} \\
\hline
Gayathiri et al. \cite{gayathiri2025c}, 2025 & CNN + DSAE & 90.9 \\
Abirami et al. \cite{abirami2025classification}, 2025 & CNN & 93.6 \\
Islam et al. \cite{islam2024improved}, 2024 & 2D CNN + LSTM & 98.47 \\
Aamir et al. \cite{aamir2024brain}, 2024 & CNN & 97 \\
Khan et al. \cite{khan2025lead}, 2025 & LEADCNN & 98.7 \\
Islam et al. \cite{islam2024improved}, 2024 & Ensemble & 98.82 \\
Proposed & CNN & \textbf{99.03} \\
\hline
\end{tabular}
\end{table}

\begin{table}[t]
\centering
\caption{Comparative analysis with Proposed CNN on Dataset 2}
\label{tab:dataset2_accuracy_sorted}
\begin{tabular}{|l|c|c|}
\hline
\textbf{Study} & \textbf{Technique} & \textbf{Accuracy} \\
\hline
Shaha et al. \cite{shaha2025mri}, 2025 & InceptionV3 & 94.12 \\
Agrawal et al. \cite{agrawal2024multifenet}, 2024 & MultiFeNet CNN & 96.4 \\
Batool et al. \cite{batool2025lightweight}, 2025 & M-CNN & 96.03 \\
Incir et al. \cite{incir2024improving}, 2024 & EfficientNetV2-M & 98.01 \\
Ilani et al. \cite{ilani2025t1}, 2025 & U-Net & 98.56 \\
Proposed & CNN & \textbf{99.28} \\
\hline
\end{tabular}
\end{table}

To further validate the performance of our proposed CNN, we investigate some latest studies that has been published in this field. Where the proposed lightweight CNN consistently outperformed several established techniques across both benchmark datasets. On Dataset 1 in Table \ref{tab:dataset1_accuracy_sorted}, it surpassed the performance of conventional CNNs and hybrid architectures such as CNN+DSAE and 2D CNN+LSTM, achieving the best classification accuracy of 99.03\%. On Dataset 2 in Table \ref{tab:dataset2_accuracy_sorted}, the proposed model achieved 99.28\% accuracy, outperforming deeper and more complex models like EfficientNetV2-M and InceptionV3, which recorded lower accuracies of 98.01\% and 94.12\%, respectively. These results demonstrate that our custom CNN, while being computationally efficient, delivers state-of-the-art performance, making it an excellent option for clinical applications with limited resources and real-time.

\section{Conclusion}

In this investigation, we suggested a lightweight custom Convolutional Neural Network (CNN) for brain tumor classification, which outperformed a number of cutting-edge pre-trained models on both Dataset 1 and Dataset 2. As shown in the experiments, our suggested architecture acquired the maximum accuracy (99.03\% and 99.29\% respectively on both Dataset 1 and 2), with significantly better precision, recall, and F1-scores in every category of tumors. This was achieved despite the proposed model having fewer parameters compared to larger pre-trained models like DenseNet201 and Xception, highlighting its efficiency. This superior performance, combined with its computational efficiency, makes our model a viable method for classifying brain tumors in settings with low resources. The model's optimization for increased computing efficiency will be the main focus of future development, exploring transfer learning to enhance performance on smaller datasets, and extending the model to multi-modal data for improved generalizability. Additionally, we aim to implement a real-time clinical version to facilitate practical use in medical diagnostics.

\vspace{12pt}

\bibliographystyle{IEEEtran}
\bibliography{ref}

\end{document}